\documentclass[12pt, onecolumn]{IEEEtran}
\usepackage{epsf}
\usepackage{graphicx}
\usepackage{amsmath} \usepackage{amssymb}

\long\def\symbolfootnote[#1]#2{\begingroup%
\def\thefootnote{\fnsymbol{footnote}}\footnote[#1]{#2}\endgroup}

\newtheorem{theorem}{Theorem}[section]

\def\dref#1{(\ref{#1})}

\def\be{\begin{equation}} \def\ee{\end{equation}}
\def\ba{\begin{array}} \def\ea{\end{array}} \def\bna{\begin{eqnarray}}
\def\ena{\end{eqnarray}}

 \def\bna{\begin{eqnarray}}
\def\ena{\end{eqnarray}} \def\dref#1{(\ref{#1})}

\begin{document}
\title{On Information-Theoretic Scaling Laws for Wireless Networks}
\author{Liang-Liang~Xie \\ \small Department of Electrical and Computer Engineering \\ \small University of Waterloo, Canada\\ \small Email: llxie@ece.uwaterloo.ca}

\maketitle

\begin{abstract}
With the analysis of the hierarchical scheme, the potential influence of the pre-constant in deriving scaling laws is exposed. It is found that a modified hierarchical scheme can achieve a throughput arbitrarily times higher than the original one, although it is still diminishingly small compared to the linear scaling. The study demonstrates the essential importance of the throughput formula itself, rather than the scaling laws consequently derived.
\end{abstract}

\section{Introduction}

Scaling-law study of the capacity of wireless networks is a retreat when the exact characterization is out of reach. Although it aims at lower goals, it opens an avenue of obtaining concrete results. Such results are asymptotic in nature, but can be very insightful especially for networks with a large number of nodes.

Consider a wireless network of $n$ nodes, where each node is an independent source and wants to send information to some other node in the network. What are the achievable rates? For this problem, the seminal work \cite{gupkum00} showed that the multi-hop operation achieves a scaling law of $\Theta(\sqrt{n})$ for the total throughput. That is, on average, each source-destination pair enjoys a rate of $\Theta(\sqrt{n})/n=\Theta(\frac{1}{\sqrt{n}})$, which unfortunately tends to zero as $n$ goes to infinity. This was not good news! It implies that no constant rate can be maintained for all source-destination pairs when the network size $n$ grows. Obviously, in order to maintain a constant rate, a linear scaling $\Theta(n)$ of the throughput has to be achieved.

Although the multi-hop operation has indeed been the focus of much protocol development, it is well known from multi-user information theory that there are many cooperation schemes that can achieve higher rates. Hence, the question remains: Is linear scaling achievable, if based on multi-user cooperations?

Recently, a hierarchical scheme based on multi-user cooperations was proposed in \cite{ozglevtse07}, where, it was shown that for any $\epsilon>0$, the scaling $\Theta(n^{1-\epsilon})$ is achievable under some network conditions. This is a significant improvement over the scaling $\Theta(\sqrt{n})$ achieved by the multi-hop operation. However, the paper \cite{ozglevtse07} cannot claim that linear scaling is achievable although $\epsilon$ can be made arbitrarily small, due to the reason that the pre-constant of the scaling is $\epsilon$-dependent and actually decreases to zero as $\epsilon$ decreases to zero.

There is a subtle difference between the scaling-law study in \cite{gupkum00} and the scaling-law study in \cite{ozglevtse07}. In \cite{gupkum00}, the pre-constant of the scaling is easy to determine due to the fixed link rate in the multi-hop operation, which does not change even when the network size grows. However, it is not so simple for networking strategies based on multi-user cooperations, which is the case in \cite{ozglevtse07}. Unfortunately, in \cite{ozglevtse07}, the pre-constant was not addressed. However, negligence of the pre-constant results in incomplete pictures, and can even lead to misleading conclusions.

In \cite{ozglevtse07}, different scaling laws were claimed for dense networks (in a fixed area) and extended networks (with a fixed density) as the number of nodes goes to infinity. However, any practical network is in a fixed area, and with a fixed density. It can either be embedded into a series of increasingly denser networks, or a series of increasingly more extended networks. Then, what can the two different scaling laws tell about the design and operation of this practical network if they are contradicting to each other? Well, the only explanation is that the scaling laws must be irrelevant to the design and operation of any practical network that lies in a fixed area and has a fixed density.

Is there anything wrong? Not really, if one takes into account the pre-constant. Consider the following simple equation:
$$
c_1 n^{\gamma_1}=c_2 n^{\gamma_2}.
$$
Obviously, for any $\gamma_1>\gamma_2$, we can always find $c_1<c_2$ for the above equation to hold for any $n$. This actually indicates that the scaling exponent $\gamma$ can be made arbitrarily large if the pre-constant $c$ is not fixed.

Therefore, without addressing the pre-constant, the scaling laws claimed in \cite{ozglevtse07} are susceptible to the ambiguity indicated above. Indeed, in \cite{ozglevtse07}, the way of improving the scaling exponent is by increasing the number of hierarchical layers $h$, such that the corresponding scaling order $\Theta(n^{\frac{h-1}{h}})$ can be arbitrarily close to linear as $\frac{h-1}{h}\rightarrow 1$. However, the unaddressed pre-constant is actually $h$-dependent, and decreases to zero as $h$ goes to infinity as demonstrated in \cite{ghaxieshe08,ghaxieshe08a}. That is, the correct and complete expression should be $c(h) n^{\frac{h-1}{h}}$, with $c(h)\rightarrow 0$ as $h\rightarrow \infty$, instead of a single $\Theta$ which cannot uncover the whole story.

 The more careful study \cite{ghaxieshe08,ghaxieshe08a} of the hierarchical scheme showed that it is not always better to choose larger $h$ for any fixed $n$. Actually, for any $n$, the optimal $h$ to choose is
\begin{equation}
\label{opth}
h^*(n)=\sqrt{\log_\beta(n/2)}
\end{equation}
where $\beta$ is a constant depending on the basic SINR (signal to interference-plus-noise ratio) in the network. This implies that any larger $h$ will result in a bigger loss in $c(h)$ compared to the gain from $n^{\frac{h-1}{h}}$. It was also shown that with the optimal choice of $h$ and the corresponding optimal cluster sizes, the maximum achievable throughput by the hierarchical scheme is
\begin{equation}
\label{Tdef}
T^*(n)=\frac{\beta R}{\sqrt{\log_\beta(n/2)}} (n/2)^{1-\frac{2}{\sqrt{\log_\beta(n/2)}}}
\end{equation}
where $R$ is another constant, also depending on the basic SINR in the network. It can be easily checked that
$$
\frac{T^*(n)}{n}\downarrow 0.
$$
That is, compared to linear scaling, the throughput achieved by the hierarchical scheme is monotonely getting worse as $n$ increases, and the average rate per source-destination pair goes to zero.

One might argue that the scaling exponent $1-\frac{2}{\sqrt{\log_\beta(n/2)}}$ in \dref{Tdef} does converge to 1 as $n\rightarrow \infty$, and thus, can be replaced by $1-\epsilon$ for arbitrarily small $\epsilon>0$, the same as the expression in \cite{ozglevtse07}. However, note that this $\epsilon$ is $n$-dependent, and smaller $\epsilon$ requires larger $n$, which in turn magnifies the importance of $\epsilon$. This is exactly why $T^*(n)$ becomes arbitrarily times worse than $n$ although the exponent does converge to 1.

But still, does this matter, if it can be claimed that any scaling of $\Theta(n^{1-\epsilon})$ is achievable for any fixed $\epsilon>0$, although the pre-constant is $\epsilon$-dependent, and diminishes to zero as $\epsilon\rightarrow 0$? Yes, it matters, for practical design and operation of wireless networks, if the scaling law studies intend to be insightful or even relevant. First, as explained above, it becomes clear that for any practical network, it is not always better to choose more hierarchical layers. More layers do increase the exponent, but also introduce more overhead when supporting the hierarchical structure. There will be some point, beyond which, the overhead overtakes the benefit of adding more layers. As a simple example, for the case where $\beta=10$, \dref{opth} shows that the optimal number of layers for a network of 20000 nodes is 2, i.e., the simplest three-phase operation in the hierarchical scheme, and the corresponding throughput is $c(2)\sqrt{20000}$, which actually is of the same order as that offered by the simple multi-hop operation. Then whether to use the hierarchical scheme or the multi-hop scheme is completely determined by the pre-constant.

Moreover, even if concentrating on the limiting behavior as $n\rightarrow \infty$, we will show in this paper that a modification of the hierarchical scheme can achieve a throughput $T_1^*(n)$ that can be arbitrarily times better than $T^*(n)$ in the sense that
$$
\frac{T_1^*(n)}{T^*(n)}\rightarrow \infty, \quad \mbox{ as } n\rightarrow \infty.
$$
Actually, a more careful evaluation shows that
$$
\frac{T_1^*(n)}{T^*(n)\log_a n}\rightarrow \infty, \quad \mbox{ for any }a>1.
$$
The potential of discovering such more superior schemes may have been ignored if one overlooked the importance of $\epsilon$ or the pre-constant.

The remainder of the paper is organized as the following. In Section II, we point out an immediate improvement that can be made on the hierarchical scheme proposed in \cite{ozglevtse07}, and introduce a modification. The throughput analysis and optimization of the modified scheme will be carried out in Section III, and will also be compared to the original scheme. In Section IV, we discuss the drawbacks associated with the notions of ``dense'' and ``extended'' networks, so artificially coined for scaling law studies, and propose a unified and direct way of addressing the real issues. Finally, some concluding remarks are presented in Section V.

\section{Clustering Multiple-Access with Relay}

We introduce a simple modification to the hierarchical scheme proposed in \cite{ozglevtse07}. The basic element in the modification is multiple-access. That is, multiple nodes want to send their independent bits to the same node simultaneously. However, instead of accomplishing this in one step, we use a hierarchical structure, where the bits are relayed via multiple levels of clusters until reaching the final destination. Before going into the details, let us first examine the scheme in \cite{ozglevtse07} to see where improvements can be made.

The network under study consists of $n$ nodes. There are $n$ source-destination pairs evenly distributed, so that each node is a source for some other node, and also is the destination of some other source. For convenience, let's call this the original S-D pair problem. In order to introduce cooperations, the network is first divided into clusters, each of $M_1$ nodes.

The basic element in the scheme in \cite{ozglevtse07} is the three-phase operation. That is, first a source node distributes its bits to the other nodes in the same cluster (different bits to different nodes); then, the source cluster sends all these bits to the destination cluster via the virtual MIMO channel; at last, all the nodes in the destination cluster send their quantized observations to the destination node. Since all nodes are sources, the first step needs to be carried out $M_1$ times for all the nodes in the source cluster, which constitute Phase 1; similarly since all nodes are destinations, the last step also needs to be carried out $M_1$ times for all the nodes in the destination cluster, which constitute Phase 3; moreover, the second step needs to be carried out $n$ times for $n$ S-D pairs, which constitute Phase 2.

Note that in each cluster of $M_1$ nodes, Phase 1 can actually be decomposed into $M_1-1$ original S-D pair problems with non-overlapping destination distributions; and similarly can Phase 3 be decomposed. It is exactly this observation which leads to the hierarchical structure proposed in \cite{ozglevtse07}, where, both Phase 1 and Phase 3 can be replaced by another three-phase operation with smaller sub-clusters of size $M_2$. Then again, the Phases 1 and 3 of the sub-clusters can be replaced by another three-phase operation with even smaller sub-sub-clusters. This process is continued, with each Phase 1 or Phase 3 being replaced by a three-phase operation with smaller clusters, and the hierarchy is built.

Our modification arises from a different perspective on Phase 1 and Phase 3. Although they can be decomposed into a sequence of the original S-D problems, they are essentially a problem where every node wants to send to every other node an independent message. From the receiver point of view, each node sees the other nodes trying to send independent messages to it via a multiple-access channel. Hence, with this new perspective, in a cluster of $M_i$ nodes, both Phase 1 and Phase 3 can be carried out by $M_i$ multiple-access operations. Since this is a task where there are multiple-accesses to all the nodes, it is convenient to name it the all-way multiple-access problem.

\begin{figure}[hbt]
\centering
\includegraphics[width=5in]{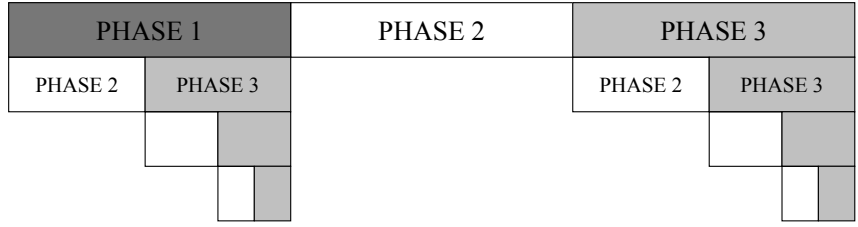}
\caption{A modified hierarchical scheme that can achieve a throughput arbitrarily higher than the original one.} \label{fig1}
\end{figure}

The advantage with this new perspective is that with cluster cooperation, the all-way multiple-access problem can be accomplished in two-steps, instead of three. That is, first the nodes in any one cluster send their bits to the destination cluster via the virtual MIMO channel; then all the nodes in the destination cluster send their quantized observations to the destination node. In other words, the first step of one node distributing its bits is not longer necessary, because now every node has something to transmit to the same destination. Correspondingly, the hierarchy proposed in \cite{ozglevtse07} can be modified as in Fig.~\ref{fig1}. Compared to the Figure 3 in \cite{ozglevtse07}, the difference is the elimination of all the Phase 1's from the hierarchy, except on the top layer, where the problem is still the original S-D pair problem, which cannot be turned into a multiple-access problem.

As stated in \cite{ozglevtse07}, the functionality of Phase 1 is for a node to distribute its bits to the other nodes in the cluster, in order to establish a virtual multi-antenna transmitter for the MIMO communication in Phase 2. However, in retrospect, since different bits are distributed to different nodes, there is essentially no mutual understanding among these nodes when they are transmitting together to the destination cluster. Therefore it may be more accurate to think of Phase 2 as a multiple-access communication with a virtual receive cluster. With this in mind, then it becomes obvious that Phase 2 can be directly carried out without the preparation of Phase 1 if the problem is already multiple-access.

The same modification with a multiple-access perspective has also appeared in \cite{ozglev08} in the context of minimizing delay. However, the authors there simply claim that the modified scheme achieves the same throughput as the original scheme in \cite{ozglevtse07}, largely due to the negligence of the $\epsilon$ as we explained in the Introduction. In next section, we will show that the modified scheme can achieve a throughput arbitrarily higher than the original scheme as the network size $n$ grows.

\section{Analysis of the Scaling Laws}

In this section, we analyze the optimal throughput achievable by the modified hierarchical scheme proposed in last section. The procedure is similar to that in \cite{ghaxieshe08,ghaxieshe08a} when analyzing the original hierarchical scheme of \cite{ozglevtse07}. It turns out that the improvement can be arbitrarily times large as the network size $n$ grows to infinity, i.e.,
$$
\frac{T_1^*(n)}{T^*(n)}\rightarrow \infty
$$
where, $T_1^*(n)$ is the optimal throughput by the modified scheme, and $T^*(n)$ is the optimal throughput by the original scheme. A more careful evaluation even shows that
$$
\frac{T_1^*(n)}{T^*(n)\log_a n}\rightarrow \infty, \quad \mbox{ for any }a>1.
$$
However, still, the average rate per S-D pair goes to zero as $n\rightarrow \infty$, i.e.,
$$
\frac{T_1^*(n)}{n}\rightarrow 0.
$$

Since the analysis procedure is similar to that in \cite{ghaxieshe08,ghaxieshe08a}, we only highlight the differences here. Note that the top layer of the hierarchy remains the same. The key issue is to determine the time needed to accomplish the all-way multiple-access problem in Phase 1 and Phase 3 of the top layer.

As defined in last section, the all-way multiple-access problem under study can be stated as the following. Consider a network of size $M_1$, where, every node wants to send $L$ bits to every other node in the network. (Different bits for different pair, i.e., totally, $M_1^2L$ bits\footnote{The accurate number should be $M_1(M_1-1)L$. However, for simplicity and without loss of much accuracy when $M_1$ is large, we use $M_1^2L$ in the calculation. This approximation won't affect the scaling order.} need to be communicated.) The question is how long it takes to accomplish the task?

We use the modified two-phase operation scheme to accomplish the task. First, we build the hierarchical structure. Divide these $M_1$ nodes into clusters of size $M_2$; then divide each cluster of $M_2$ nodes into smaller clusters of size $M_3$; continue this process $h-2$ times for some $h\geq 2$, and finally we obtain clusters of size $M_{h-1}$. We will determine the optimal value of $h$ to stop, i.e., the optimal number of hierarchical layers, and also the optimal cluster sizes $M_2$, $M_3$, $\ldots$, $M_{h-1}$ in the sequel.

Obviously, the number of time slots needed to accomplish the all-way multiple-access problem with the above hierarchical structure depends on the parameters $h$, $M_1$, $M_2$, $\ldots$, $M_{h-1}$, $L$, and therefore, is denoted by
\begin{equation}\label{delay}
D_{h-1}(M_1,M_2,\ldots,M_{h-1},L).
\end{equation}
We'll use a recurrence relation to determine \dref{delay}. First, note that the all-way multiple-access problem of the network of size $M_1$ is accomplished in two phases: Phase 2 and Phase 3, with clusters of size $M_2$. The number of time slots needed for Phase 2 is simply $\frac{M_1}{M_2}2M_1\frac{L}{R}$, as calculated in \cite{ghaxieshe08,ghaxieshe08a}, where $R$ is the basic rate. In Phase 3, it is again the all-way multiple-access problem for networks of smaller size $M_2$, but now, with $L\frac{Q}{R}\frac{M_1}{M_2}$ bits to be communicated between each pair of nodes. Hence, we have the relation
$$
\begin{array}{l}
D_{h-1}(M_1,M_2,\ldots,M_{h-1},L) \\ [-2mm] \\
=\frac{M_1}{M_2}2M_1\frac{L}{R}+4 D_{h-2}(M_2,\ldots,M_{h-1},L\frac{Q}{R} \frac{M_1}{M_2})
\end{array}
$$
where the multiplier $4$ is needed for time-sharing between neighboring clusters to avoid excessive interference.

In turn, we have the following relation
$$
\begin{array}{l}
D_{h-2}(M_2,\ldots,M_{h-1},L\frac{Q}{R} \frac{M_1}{M_2}) \\ [-2mm] \\
=\frac{M_2}{M_3}2M_2\frac{L}{R}\frac{Q}{R}\frac{M_1}{M_2}+4 D_{h-3}(M_3,\ldots,M_{h-1},L\frac{Q}{R} \frac{M_1}{M_2}\frac{Q}{R}\frac{M_2}{M_3})
\end{array}
$$
and similar recursive relations for $D_{h-3}$, $D_{h-4}$ and so on.
Hence, recursively,
$$
\begin{array}{l}
D_{h-1}(M_1,M_2,\ldots,M_{h-1},L) \\ [-2mm] \\
=\frac{M_1}{M_2}2M_1\frac{L}{R} \\ [-2mm] \\ \quad +4\frac{M_2}{M_3}2M_2\frac{L}{R}\frac{Q}{R}\frac{M_1}{M_2} \\ [-2mm] \\ \quad
+4^2\frac{M_3}{M_4}2M_3\frac{L}{R}(\frac{Q}{R})^2\frac{M_1}{M_3} \\ [-2mm] \\ \quad
+ \cdots \\ [-2mm] \\ \quad
+4^{h-3}\frac{M_{h-2}}{M_{h-1}}2M_{h-2}\frac{L}{R}(\frac{Q}{R})^{h-3}\frac{M_1}{M_{h-2}} \\ [-2mm] \\ \quad
+4^{h-2}D_1(M_{h-1},L(\frac{Q}{R})^{h-2}\frac{M_1}{M_{h-1}}).
\end{array}
$$

For the smallest clusters of size $M_{h-1}$, the all-way multiple-access problem is accomplished directly without the two-phase operation, and thus
$$
D_1(M_{h-1},L(\frac{Q}{R})^{h-2}\frac{M_1}{M_{h-1}})=\frac{L}{R}(\frac{Q}{R})^{h-2} \frac{M_1}{M_{h-1}}M_{h-1}^2.
$$
Therefore, letting $c=4\frac{Q}{R}$,
$$
D_{h-1}(M_1,M_2,\ldots,M_{h-1},L)
=2M_1\frac{L}{R}\left(\frac{M_1}{M_2}+c\frac{M_2}{M_3}+c^2\frac{M_3}{M_4}+ \cdots+c^{h-3}\frac{M_{h-2}}{M_{h-1}}+c^{h-2}\frac{M_{h-1}}{2}\right).
$$

For any fixed $M_1$, to minimize the sum in the parenthesis above, noting that the product of all those terms is
$$
c^{1+2+\cdots+(h-2)}\cdot \frac{M_1}{2}=c^{\frac{(h-1)(h-2)}{2}}\cdot \frac{M_1}{2},
$$
obviously, the optimal choice is that every term equals to
$$
\left(c^{\frac{(h-1)(h-2)}{2}}\cdot \frac{M_1}{2}\right)^{\frac{1}{h-1}}=c^{\frac{h-2}{2}}(\frac{M_1}{2})^{\frac{1}{h-1}}.
$$
This leads to the optimal choices of cluster sizes:
\begin{equation}
\label{clustersize}
M_i=2c^{-\frac{(i-1)(h-i)}{2}}(\frac{M_1}{2})^{\frac{h-i}{h-1}},\quad \quad 2\leq i\leq h-1
\end{equation}
and the minimum number of time slots:
$$
D^*_{h-1}(M_1,M_2,\ldots,M_{h-1},L)= 2M_1\frac{L}{R}(h-1)c^{\frac{h-2}{2}}(\frac{M_1}{2})^{\frac{1}{h-1}}.
$$

Therefore, on the top layer, the number of time slots needed for Phase 1 is
$$
4\times 2M_1\frac{L}{R}(h-1)c^{\frac{h-2}{2}}(\frac{M_1}{2})^{\frac{1}{h-1}};
$$
the number of time slots needed for Phase 3 is
$$
4\times 2M_1\frac{L}{R}(h-1)c^{\frac{h-2}{2}}(\frac{M_1}{2})^{\frac{1}{h-1}}\frac{Q}{R};
$$
and the number of time slots needed for Phase 2 is still $2n\frac{L}{R}$, the same as the original scheme.
After these time slots, the number of bits transported for each S-D pair is $M_1L$, and the total number of bits transported in the whole network is $nM_1L$. Therefore, the throughput is calculated as
$$
\frac{nM_1L}{16\frac{L}{R}(h-1)(1+\frac{Q}{R}) c^{\frac{h-2}{2}}(\frac{M_1}{2})^{\frac{h}{h-1}}+2n\frac{L}{R}}=:f(M_1).
$$
It is easy to find the optimal choice of $M_1$ by setting $f'(M_1)=0$, and we have
\begin{equation}
\label{clustersize1}
n=8(1+\frac{Q}{R})c^{\frac{h-2}{2}}(\frac{M_1}{2})^{\frac{h}{h-1}} \quad \mbox{ or equivalently, } \quad M_1=2\left[8(1+\frac{Q}{R})c^{\frac{h-2}{2}} \right]^{-\frac{h-1}{h}} n^{\frac{h-1}{h}}
\end{equation}
and the corresponding throughput
$$
T^{opt}_h(n)=\frac{R}{h(1+{R}/{Q})^{\frac{h-1}{h}} (4Q/R)^{\frac{h-1}{2}}}(n/2)^{\frac{h-1}{h}}.
$$

For any fixed $n$, we can find the optimal $h$ to maximize $T^{opt}_h(n)$ by setting
$$
\frac{dT^{opt}_h(n)}{dh}=0.
$$
This leads to
$$
h^2 \ln(2\sqrt{Q/R})+h-[\ln(n/2)-\ln(1+R/Q)]=0.
$$
Hence, the optimal number of layers to choose is
$$
h^*=\frac{\sqrt{1+4\ln(2\sqrt{Q/R})[\ln(n/2)-\ln(1+R/Q)]}-1}{2\ln(2\sqrt{Q/R})}.
$$
Similarly as in \cite{ghaxieshe08,ghaxieshe08a}, in order to obtain a simple formula, we use the approximation
\begin{equation}
\label{hdef}
h^*=\frac{\sqrt{4\ln(2\sqrt{Q/R})\ln(n/2)}}{2\ln(2\sqrt{Q/R})}
\end{equation}
which is very accurate for large $n$. Letting $\beta_1=2\sqrt{Q/R}$, we have
\begin{equation}
\label{h*def}
h^*=\sqrt{\frac{\ln(n/2)}{\ln(2\sqrt{Q/R})}}=\sqrt{\log_{\beta_1}(n/2)}.
\end{equation}
Note that
$$
\beta_1^h=\beta_1^{\log_{\beta_1}(n/2)\frac{h} {\log_{\beta_1}(n/2)}}=(n/2)^{\frac{h}{{\log_{\beta_1}(n/2)}}}.
$$
Therefore,
\begin{eqnarray}
T^{opt}_h(n)&=&\frac{R}{h(1+{R}/{Q})^{\frac{h-1}{h}} (4Q/R)^{\frac{h-1}{2}}}(n/2)^{\frac{h-1}{h}}  \nonumber  \\
&=& \frac{\beta_1 R}{h (1+R/Q)^{\frac{h-1}{h}} \beta_1^h} (n/2)^{\frac{h-1}{h}} \nonumber \\
&=& \frac{\beta_1 R}{h (1+R/Q)^{\frac{h-1}{h}} } (n/2)^{1-\frac{1}{h}-\frac{h}{\log_{\beta_1}(n/2)}} \label{Thdef}
\end{eqnarray}
where letting $h=h^*=\sqrt{\log_{\beta_1}(n/2)}$, we have the optimal throughput
\begin{equation}
\label{T1def}
T^*_1(n)=\frac{\beta_1 R}{c_n\sqrt{\log_{\beta_1}(n/2)}}(n/2)^{1-\frac{2}{\sqrt{\log_{\beta_1}(n/2)}}}
\end{equation}
where
$$
c_n=(1+R/Q)^{1-\frac{1}{\sqrt{\log_{\beta_1}(n/2)}}}\rightarrow (1+R/Q), \quad \mbox{ as }n \rightarrow \infty.
$$
Obviously, \dref{T1def} is very accurate for large $n$, although we made some approximation in \dref{hdef} and $h^*$ should always be an integer.

Hence, we arrive at the following theorem.
\begin{theorem}
With the modified hierarchical scheme, by choosing the optimal number of layers as \dref{h*def} and the corresponding optimal cluster sizes as \dref{clustersize} and \dref{clustersize1}, the optimal throughput is given by \dref{T1def}.
\end{theorem}

Without the approximation \dref{hdef}, we can also obtain an exact upper bound of the throughput as the following. By \dref{Thdef},
\begin{eqnarray*}
T^{opt}_h(n)&\leq& {\beta_1 R}  (n/2)^{1-\frac{1}{h}-\frac{h}{\log_{\beta_1}(n/2)}} \\
&\leq& {\beta_1 R}  (n/2)^{1-\frac{2}{\sqrt{\log_{\beta_1}(n/2)}}}
\end{eqnarray*}
where, in the last inequality, ``$=$'' holds if $h=\sqrt{\log_{\beta_1}(n/2)}$. It is easy to check that with the modified hierarchical scheme, the average rate per S-D pair still goes to zero as
\begin{eqnarray*}
\frac{(n/2)^{1-\frac{2}{\sqrt{\log_{\beta_1}(n/2)}}}}{n} &=& \frac{1}{2} (n/2)^{-\frac{2}{\sqrt{\log_{\beta_1}(n/2)}}}   \\
&=& \frac{1}{2} \left(\beta_1^{\log_{\beta_1}(n/2)}\right)^{-\frac{2}{\sqrt{\log_{\beta_1}(n/2)}}} \\
&=& \frac{1}{2} \beta_1^{-2\sqrt{\log_{\beta_1}(n/2)}}  \\ [-3mm]  \\
&\rightarrow& 0.
\end{eqnarray*}

However, the modified scheme can be arbitrarily times better than the original one in \cite{ozglevtse07}, as can be checked with
\begin{eqnarray*}
\frac{T^*_1(n)}{T^*(n)} &=& \frac{\beta_1 R}{c_n\sqrt{\log_{\beta_1}(n/2)}} (n/2)^{1-\frac{2}{\sqrt{\log_{\beta_1}(n/2)}}}\Bigg/\frac{\beta R}{\sqrt{\log_{\beta}(n/2)}} (n/2)^{1-\frac{2}{\sqrt{\log_{\beta}(n/2)}}} \\ \\
&=&\frac{\beta_1\sqrt{\log_\beta \beta_1}}{c_n\beta}(n/2)^{\frac{2}{\sqrt{\log_{\beta}(n/2)}}(1-\sqrt{\log_\beta \beta_1})} \\ \\
&=&\frac{\beta_1\sqrt{\log_\beta \beta_1}}{c_n\beta}\beta^{{2}(1-\sqrt{\log_\beta \beta_1}){\sqrt{\log_{\beta}(n/2)}}}  \\ [-2mm]  \\
&\rightarrow& \infty,
\end{eqnarray*}
where, $T^*(n)$ is the optimal throughput of the original scheme as calculated in \cite{ghaxieshe08,ghaxieshe08a} with $\beta=2\sqrt{1+Q/R}>\beta_1$, and thus, $\log_{\beta}\beta_1<1$. Actually, we can show an even stronger result that
$$
\frac{T_1^*(n)}{T^*(n)\log_a n}\rightarrow \infty, \quad \mbox{ for any }a>1,
$$
since
$$
\frac{2\left(1-\displaystyle\sqrt{\log_\beta \beta_1}\,\right)\sqrt{\log_\beta(n/2)}}{\log_\beta\log_a n}\rightarrow \infty.
$$

\section{Dense or Sparse Networks?}

The analysis in last section has assumed a fixed basic rate $R$ in order to focus on the scaling in terms of $n$. While this is the case under some channel gain model for the so-called dense networks, where networks are confined in a fixed area even as the number of nodes grows to infinity, it is not so easy to maintain a fixed basic rate for networks with growing areas, due to the power path loss.

Therefore, when addressing the so-called extended networks, where the node density is fixed while the area grows proportionally to the number of nodes, \cite{ozglevtse07} proposed the trick of concentrating the total transmission power into a small portion of the total transmission time to compensate for the path loss, so that during that  portion, the received SINR is maintained at a specific level. Then, with the following power path loss model:
$$
P_r=P_t/d^\alpha,
$$
i.e., the received power $P_r$ depends on the transmitted power $P_t$ via the transmitter-receiver distance $d$ and the path-loss exponent $\alpha$,
the scaling for extended networks readily follows by multiplying all the results above with the factor $n^{1-{\alpha}/{2}}$. Namely, to compensate for the power path loss, the transmitted power needs be $d^\alpha$ times larger, i.e., $(\sqrt{n})^\alpha$ times larger considering long-hop distances in an area proportional to $n$. Since the required power level is $P/n$ for dense networks, the hierarchical scheme can only be operated in $n^{1-\alpha/2}$ portion of the time for extended networks to satisfy the total power constraint, which leads to the multiplication of the same factor to all the scaling law results obtained previously.

More generally, as pointed out in \cite{ghaxieshe08,ghaxieshe08a}, the same trick can be played on networks with any area other than either fixed, or linear growing. That is, a network with area $A$ is distinguished into two categories based on whether
\begin{equation}
\label{an}
A^{\alpha/2}\leq n.
\end{equation}
In the case where $A^{\alpha/2}\leq n$, the basic SINR can be maintained all the time, and the power-concentration trick is not needed; In the other case where $A^{\alpha/2}> n$,  the power-concentration trick is needed to maintain the basic SINR for $n/A^{\alpha/2}$ portion of the time, and all the results correspondingly need to be multiplied by the same factor. For example, the formula \dref{T1def} should be modified as
\begin{equation}
\label{T1def1}
T^*_1(n,A)=\min\left\{1, \frac{n}{A^{\alpha/2}} \right\} \frac{\beta_1 R}{c_n\sqrt{\log_{\beta_1}(n/2)}}(n/2)^{1-\frac{2}{\sqrt{\log_{\beta_1}(n/2)}}}.
\end{equation}
In \cite{ghaxieshe08,ghaxieshe08a}, these two categories are respectively named as ``dense'' and ``sparse'' networks. Note that this is a notion that can be readily clarified on any specific network based on the relation between the area $A$ and the number of nodes $n$, different from the previous notion of ``dense'' and ``extended'' networks that is undetermined for any single network. However, one has to realize that this new notion is largely a consequence of the hierarchical scheme, and is also related to the path-loss exponent.

A recent work \cite{ozgjohtselev08} proposes to address the intermediate regime between dense and extended networks by introducing a more general pattern of area scaling as
\begin{equation}
\label{asc}
A=n^\nu
\end{equation}
where $\nu$ is a real number, with $\nu=0$ corresponding to dense networks, and $\nu=1$ corresponding to extended networks. Although it seems more general by the flexibility of choosing different values for $\nu$, it is still artificial to make the network area scale according to the pattern \dref{asc}. The ambiguity of determining the right embedding process for any specific network still remains, as we pointed out in the Introduction.

Above all, the motivation of studying the capacity of wireless networks is clear: To provide insight and guidance on the practical design and operation of such networks. After the explicit determination of the pre-constant, it is clear that the throughput formulas such as \dref{T1def1} hold for any finite number $n$, and the scaling laws are just a consequence of letting $n\rightarrow \infty$. If the practical problem under study is a specific network with a specific area and a specific number of nodes, then obviously, it is more natural and insightful to apply the formula \dref{T1def1} directly rather than to consult the scaling laws thereafter derived. So probably, one should not be concerned with the scaling laws so much as the exact throughput formula itself.

Note that in the formula \dref{T1def1}, the parameters $R$, $Q$, $\beta_1$ also affect the throughput, and they are determined by the basic SINR, which in turn, is determined by the long-hop path loss. Therefore, for the flexibility of selecting different basic SINRs, the criterion \dref{an} should be modified as
\begin{equation}
\label{an2}
A^{\alpha/2}\leq c_0n,
\end{equation}
and the corresponding optimal throughput is modified as
\begin{equation}
\label{T1def2}
T^*_1(n,A)=\min\left\{1, \frac{c_0 n}{A^{\alpha/2}} \right\}\frac{\beta_1 R}{c_n\sqrt{\log_{\beta_1}(n/2)}}(n/2)^{1-\frac{2}{\sqrt{\log_{\beta_1}(n/2)}}}
\end{equation}
where, $c_0$ is a constant, chosen to set the threshold of the basic SINR, and thus the values of the parameters $R$, $Q$, $\beta_1$. Generally, smaller $c_0$ leads to larger basic rate $R$; however, smaller $c_0$ may also make the condition \dref{an2} unsatisfied, and thus lead to the scale-down factor ${c_0 n}/{A^{\alpha/2}}$ in \dref{T1def2}, as a result of the power concentration trick. Hence, there is a basic tradeoff in choosing $c_0$ when maximizing \dref{T1def2}. Apparently, the afore-mentioned notions of ``dense'' and ``sparse'' networks derived via the criterion \dref{an} are rather arbitrary and scheme-dependent than fundamental.

In summary, \dref{T1def2} presents the optimal throughput achievable by the modified hierarchical scheme for a network of $n$ nodes and area $A$. This is all we need to know. Based on this, all kinds of scaling laws can be derived by setting different limits. Now the question is really how good is \dref{T1def2}, for any possible values of $n$ and $A$, not just when $n\rightarrow \infty$. We have presented a simple example in the Introduction showing that this is a question even when only compared to the multi-hop scheme. In general, we note that the upper bounds obtained in \cite{xiekum04,xiekum06} apply to any finite network with specific $n$ and $A$, and in fact encompass more general traffic patterns with the criterion of transport capacity, which allows unequal rates and uneven S-D distributions.

\section{Conclusion}

Caution on the pre-constant is needed when deriving scaling laws for wireless networks, especially with multi-user cooperation schemes where the overhead may not be negligible. Based on explicit analysis of the pre-constant, we have shown that a modified hierarchical scheme can achieve a throughput arbitrarily times higher than the original one, although it is still diminishingly lower compared to the linear scaling. This leaves the question open whether it is possible to maintain a constant rate between each S-D pair when the number of nodes grows to infinity.

On the other hand, rather than the scaling laws, we have demonstrated the pivotal importance of the throughput formula itself as a function of the network parameters. We emphasize that all scaling laws can be derived from this formula, and more importantly, it is this formula that is directly related to practice.

\end{document}